\documentclass[preprints,article,accept,pdftex,moreauthors]{Definitions/mdpi} 
\firstpage{1} 
\pubvolume{1}
\issuenum{1}
\articlenumber{0}
\pubyear{2023}
\copyrightyear{2023}
\datereceived{} 
\dateaccepted{} 
\datepublished{} 
\hreflink{https://doi.org/} 

\usepackage{todonotes}

\usepackage{comment}
\usepackage{graphicx}
\usepackage[normalem]{ulem}

\usepackage[ruled, lined, linesnumbered, commentsnumbered, longend]{algorithm2e}

\Title{Periodicity Intensity Reveals Insights into Time Series Data: Three Use Cases}

\TitleCitation{Periodicity Intensity Reveals Insights into Time Series Data: Three Use Cases}

\Author{Alan F. Smeaton $^{1,*}$\orcidA{} and Feiyan Hu $^{1}$\orcidB{}}

\AuthorNames{Alan F. Smeaton and Feiyan Hu}

\AuthorCitation{Smeaton, Alan F.;  Hu, F.}

\address{%
$^{1}$ \quad Insight Centre for Data Analytics, Dublin City University, Glasnevin, Dublin 9, Ireland; alan.smeaton@dcu.ie}

\corres{Correspondence: alan.smeaton@dcu.ie}

\abstract{
{Periodic phenomena} are oscillating signals found in many naturally-occurring time series. A periodogram can be used to measure the intensities of  oscillations at different frequencies over an entire time series but sometimes we are interested in measuring how periodicity intensity at a specific frequency varies throughout the time series. This can be done by calculating periodicity intensity within a  window  {then sliding 
 and recalculating the intensity for the} window, giving an indication of how periodicity intensity at a specific frequency changes throughout the series.
We illustrate three applications of this the first of which is movements of a herd of new-born calves where we show how intensity of the 24h periodicity increases and decreases synchronously across the herd. 
We also show how changes in 24h periodicity intensity of activities detected from in-home sensors can be indicative of overall wellness. We illustrate this on several weeks of sensor data gathered from each of the homes of 23 older adults.
Our third application is the intensity of 7-day periodicity of hundreds of University students accessing online resources from a virtual learning environment (VLE) and how the regularity of their {weekly} learning behaviours changes throughout a teaching semester.
The paper demonstrates how periodicity intensity reveals insights into time series data not visible using other forms of analysis.}

\newcommand{\mycomment}[1]{}
\begin{document}

\def\x{{\mathbf x}}
\def\L{{\cal L}}



\section{Introduction}

A common application of data analytics to data streams or  time series is  detecting anomalous behaviour or outliers which deviate in some way from normal behaviour, where normal behaviour is defined by the recurring patterns within {the} data stream.
As noted by \citeauthor{a15110429} in \cite{a15110429} this is useful for applications like  fraud detection, network intrusion detection, medical diagnosis, video surveillance,  fault diagnosis and more.
By definition, outlier detection detects baseline patterns within  data and then detects and alert us to outliers or {to} subtle variations outside those baseline patterns.

Periodicity  is defined as the characteristic or tendency for some pattern to recur at regular intervals within a data stream. The term is used to describe the regularity  of phenomena which reoccur throughout nature as normal behaviour and from which deviations are characterised as outliers. 
Such temporal recurrence is an important property associated with complex systems and the study of such temporal recurrence is useful for gaining insights into the behaviour of {the}  systems \cite{10.1093/cz/zoz016}.

There is regular periodicity in the rotation of the earth, the lunar cycle, the earth{'s revolving} around the sun, the seasons, the tides, and the rhythm of our daily 24-hour cycle.  
Humans, and most other living creatures, have a strong circadian rhythm, the 24-hour cycle which brings routine and structure to our lives.  Chronobiology is the field within life sciences that explores how this 24-hour cycle influences the health, wellness and the behaviour of living organisms \cite{kuhlman2018introduction}. 24-hour periodicities are observed in many human behaviours as we strive for homeostasis –- equilibrium in our physiology and stability in the face of external and environmental changes \cite{cohen2010wellness}.  It is known that when we live our lives and behave in ways that map strongly to the periodicity of the circadian clock, then that is a sign of better health and improved overall wellness \cite{farhud2018circadian}.

One approach to detecting  recurrences in data streams is visualising them via  recurrence plots (RPs) \cite{marwan2007recurrence}. These are visualisations of a matrix in which elements of the matrix correspond to  instances at which there is a recurrence in a dynamic system.  RPs operate by {first} performing a cluster analysis {on the data} and then building graphical patterns to illustrate  recurrences in the data. {RPs} have been used in many applications for example fault diagnosis of machinery \cite{9427531} and {in} image transformations \cite{a14120349}.

Recurrence plots are useful for providing  analysis  of short sequences with a limited amount of recurrence, but for longer time series with many iterations of recurrences they are designed to hide the volume of iterations and to mask  variations that may occur among {the} iterations.
Sometimes when we look for {and find} patterns in time series data  we want to measure {those} patterns and to compare the patterns over time or {even} to compare them to other recurring patterns{. This goes far beyond  just using} the patterns to detect outliers.
If we could measure the patterns and how strong is their recurrence at different points  throughout {a} dataset then this might reveal insights into the patterns and  deeper insights into the data.

In previous work we  developed a technique for calculating and visualising the strength of periodicity at a given frequency, throughout a time series. The technique was applied to the sleeping patterns of US war veterans who continuously wore wrist-worn accelerometers for 90 consecutive days each and we found that the 24h periodicity strength varied significantly across participants \cite{buman2016behavioral}. 
That work also found that high levels of periodicity indicating regular sleep and movement patterns were associated with lower measures of LDL-cholesterol, triglycerides and hs-CRP as well as improved health-related quality of life measures \cite{buman2016behavioral}. 
The same technique for visualising periodicity strength over time was used in an analysis of lifelog data defined as the automatic digital recording of everyday activities, and it provided insights into shifts in underlying behaviour of the subjects  \cite{hu2016periodicity}. 

While \citeauthor{chegini2020new} \cite{chegini2020new} used the same term  ``periodicity intensity'' in their work on the diagnosis of faults in machine bearings, their work defines periodicity intensity as the ratio of the energy of the local maxima  to the total energy in the autocorrelation of vibration signals in faulty machinery and that is very different to the definition of periodicity used here.

{Our research hypothesis is that computing a visualisation of the strength of 24-hour periodicity for a range of longitudinal time series data will reveal insights into the data not visible using other forms of analysis.}
In this paper we apply the calculation of periodicity strength to three different use cases and show how, in each case, {deeper} insights into the data that are revealed by the analysis. The paper's importance lies in the relative simplicity of the technique and of the computation{,} and the powerful insights into time series data that it provides. The technique can be used on any data series where underlying temporal patterns reoccur and where the strength of those patterns throughout the data is important to {determine}.

\section{Materials and Methods}

\subsection{Calculating Periodicity Intensity Using Time-lagged Overlapping Windows}

A periodogram \cite{bartlett1950periodogram} is a well established technique for exploring the magnitudes of recurring patterns at different frequencies in a time series of data.  A periodogram uses samples of {a time series of} data to measure the amplitude vs. the frequency characteristics of a{n underlying} continuous function and thus {it} reveals the spectral power density of the underlying function.  If the function is sampled unevenly or there is missing data from the sample, then the Lomb–Scargle periodogram is a well-known algorithm which can cope with such unevenly sampled data \cite{VanderPlas_2018}.  

The Lomb–Scargle periodogram has been used widely within those domains where there is an interest in exploring both the frequencies and  magnitudes of periodicity, such as data from astronomical observations.  In the work here we are interested in the magnitude or amplitude of the data samples only at a known frequency {such as the 24-hour circadian rhythm}. Software implementations of the Lomb-Scargle periodogram are readily  available across multiple platforms including MatLab and in Python.

\begin{figure}[H]
\includegraphics[width=\linewidth]{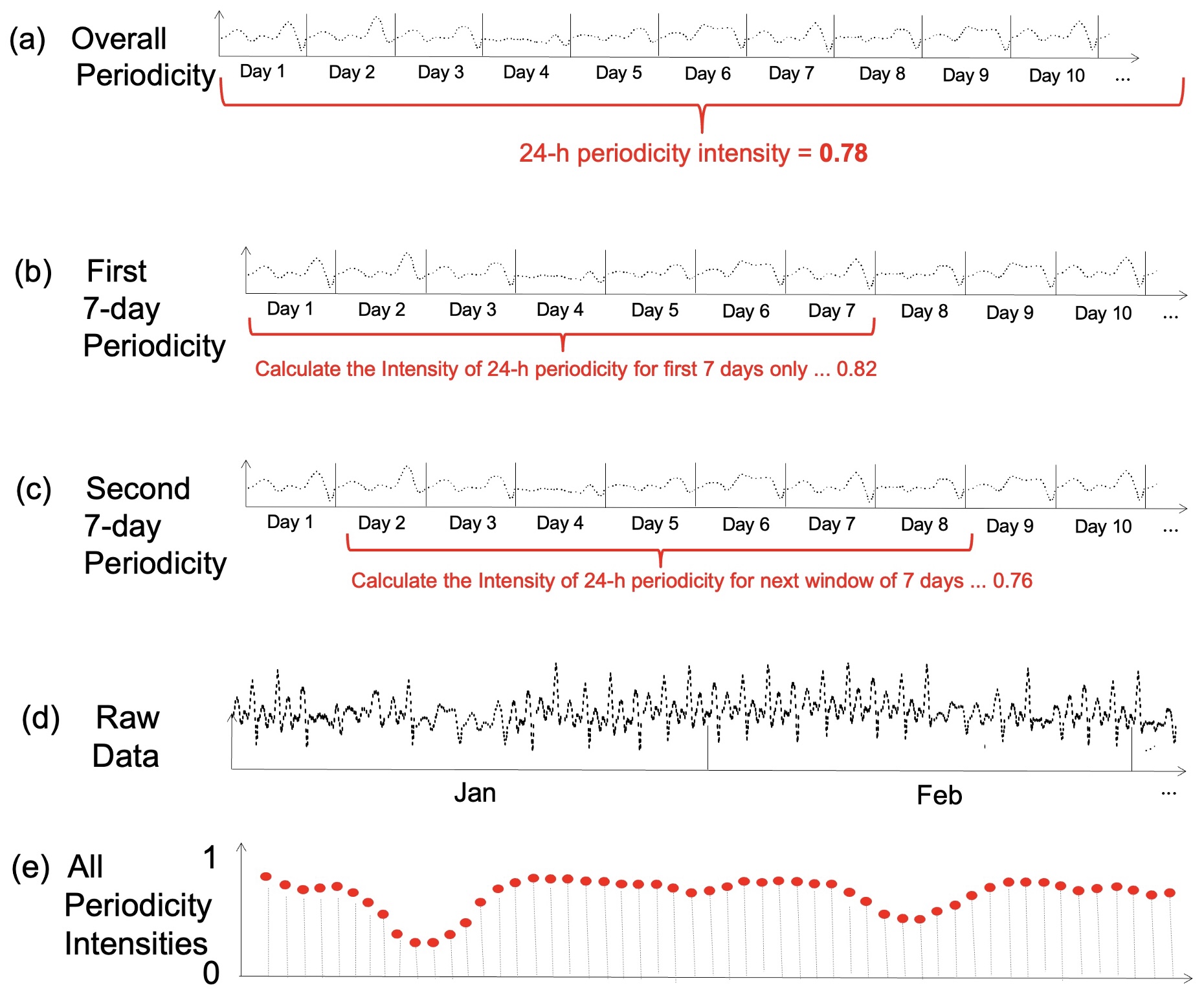}
\caption{Schematic demonstrating the calculation of periodicity intensity throughout a time series using time-lagged overlapping windows.\label{fig:periodicity}}
\end{figure}

To illustrate how we calculate the magnitude of the periodicity changes at a given frequency throughout a time series 
 using time-lagged overlapping windows, we {will describe this using}  the schematic presented in Figure~\ref{fig:periodicity}.

Figure~\ref{fig:periodicity}(a) shows a plot of the first 10 days of a longer (synthetic) series of data points and we can see from the visualisation that there is a noticeable  recurring daily pattern, i.e. when the frequency is 24 hours, although if we examine it carefully we see that each day is slightly different {from the others}.  If we compute the magnitude of this 24-hour periodicity over the whole time series it {has} a value of, {let us say,} 0.78 on a scale of 0 to 1.

While such an analysis has  benefit, if we have a long period of data (e.g. weeks, months or years{,} then calculating a single number (0.78 in the diagram) for the 24-hour periodicity over the whole duration will hide any variation{al} details within the {data}. Instead, we calculate the 24-hour periodicity {from} a window of the first 7 days only as shown in Figure~\ref{fig:periodicity}(b) and this generates a value of, say, 0.82. We then shift the 7-day window forward in time by {an interval of}, say, 1 whole day and re-calculate the 24-hour periodicity as shown in Figure~\ref{fig:periodicity}(c) and this generates a value of, say, 0.76.  We repeat this process {of shifting the 7-day window and recalculating} until we reach the end of the time series. 

Three important parameters in this {process} are (1) the frequency of the recurring pattern {$Fo$}, often 24 hours if we are interested in circadian rhythms but not always so, (2) the size of the window in which the periodicity is calculated $m$ and (3) the overlap between {the} windows $l$ also referred to as stride length, {namely} the amount by which a window is shifted before calculating a new value for {a } periodicity {window}.

For a series of $N$ data points such as that shown in Figure~\ref{fig:periodicity}(d) and a window size of {$m$} with a  shifting or stride {length} of {$l$} for each periodicity re-calculation, this generates a  time series of {$\lfloor \frac{N - m + 1}{l}\rfloor$} data points as  in Figure~\ref{fig:periodicity}(e){. This } captures  gradual behaviour changes in the regularity of the data sampled over time {rather than}  instances of particularly abnormal data values.
{Periodicity intensity graphs are just one of many possible forms of analysis of time series data but differs from each of the others. While we do perform a spectral analysis \cite{10.1145/380752.380859} of the data, this is not carried out across a full spectrum. Instead we focus on 24-hour circadian periods because previous studies indicate that circadian periodicity carries most of the energy in the longitudinal accelerometer data \cite{montaruli2021biological}. Likewise we do not map the time series data into another space like an embedded space, as has been done in \cite{ho2019using} with protein sequences.  Discrete wavelet transforms \cite{rhif2019wavelet} can be used with time series data for noise filtering and data reduction, and specifically for detecting  singularities like outliers \cite{10.1145/1883612.1883613} but this is different to our use case where we are looking to identify for long term behaviour changes.   Hilbert transforms (empirical mode decomposition) \cite{shukla2009empirical} are often used with time series for gathering information about frequency characteristics like amplitude and phase but not long-term shifts in time series characteristics. The zero crossing rate (ZCR) \cite{lartillot2007matlab} is used extensively in speech and music audio processing and corresponds to the rate at which a signal transitions from positive to negative or negative to positive  while  historical volatility \cite{hong2017general} is a statistical measure from economics and finance measuring the degree of variation of adjacent values in a time series. Each of these latter forms of analysis measure the dynamic changes in a time series but do not give the longitudinal changes in 24-hour periodicity intensity which our measure does.}

{These signal processing techniques are used in a wide variety of applications such as data augmentation to support deep learning \cite{Abay10.1007/978-3-030-61401-0_65} and in \cite{Altafs22052012} where time, frequency and spectral power domain feature vectors derived from vibration and acoustic emissions from bearings in machinery in order to detect faults. Empirical Mode Decomposition (EMD) is a technique used to decompose a signal into useful components similar to analysis methods like Fourier transforms and wavelets and it is particularly useful in exploratory analyses of data. The principal behind EMD is to split an original time series into a set of smooth curves or intrinsic mode functions. In the case of fault diagnosis it is typically used  to decompose a vibration signal into components used to train classifiers, usually with high accuracy.  However EMD is known to be sensitive to noise as well as requiring a manual selection of parameters \cite{cohen2014analyzing} to support its exploratory nature. Zerocross Density Decomposition (ZCD) is a more recently developed technique \cite{Sidekerskiene2020} which is based on the concept of zero-crossings and is computationally more efficient than EMD. However EMD is more effective than DD in terms of accuracy and robustness when dealing with non-stationary signals \cite{uzunouglu2018comparative}.
Our algorithm does not split a time series into constituents as EMD or ZCD do and our method is less sensitive to noise because we use the Lomb-Scargle periodogram \cite{VanderPlas_2018} to calculate periodicity, an algorithm which tolerates irregular sampling and missing data.
}

\SetKwFunction{len}{len}
\SetKwFunction{fft}{FFT}
\SetKwFunction{ls}{Lomb-Scargle}
\SetKwFunction{intfunc}{IntensityFunc}
\SetKwFunction{select}{Select}

{Algorithm~\ref{algo:perio_int} demonstrates the method to compute periodicity intensity. We assume we already know the optimal window size $m$, stride length $l$ and the frequency $Fo$ that we have identified and are interested in. In this case studies in this paper $Fo$ can either be circadian (24-hour) or weekly periodicity. $X_w$ is the windowed input data and $S_f$ is the periodogram computed with Fast Fourier Transform. \intfunc computes intensity by taking the periodogram of a windowed/local time series data and one focused frequency, $Fo$. We aggregate energy carried by all close frequencies that are around $Fo$ in case of energy leakage.}
\begin{algorithm}[h]
    \SetKwInOut{KwInp}{Input}
    \SetKwInOut{KwOut}{Output}

    \KwInp{A time series data $x_{n} \in X$ Where $n = 1,2,3 \dots N$ with $N$ data points. Optimal window size $m$ and  stride length $l$. $Fo \in$ \{$\frac{1}{24h}, \frac{1}{168h}$\}}
    \KwOut{A time series data $y_t \in Y$ where $t = 1, 2, 3, \dots \lfloor \frac{N - m + 1}{l}\rfloor$.}    

    \For{$t \leftarrow 0$ \KwTo $\lfloor \frac{N - m + 1}{l}\rfloor$}{
        $X_w = x_{k \times l} \dots x_{k \times l + m}$ 
        
        $S_f$ = \fft ($X_w$)
        
        $y_{t}$ = \intfunc($Fo$,  $S_f$) \textbf{where} 
        
        In this work \intfunc $= \sum_{f \in \hat{Fo}} S_f$ where $\hat{Fo}$ is the close neighbours of $Fo$ in terms of frequency satisfying $|\frac{1}{Fo} - \frac{1}{\hat{Fo}}| < 0.01$.
        
        \intfunc can also be the same form as defined in \cite{hu2016periodicity}

    }
    \caption{{Compute Periodicity Intensity}}
    \label{algo:perio_int}
    
\end{algorithm}

\subsection{Use Cases for Calculating Periodicity Intensity}

\subsubsection{Periodicity Intensity in Sensor Data from the Homes of Older Adults}

The NEX project \cite{info:doi/10.2196/35277}  developed an integrated IoT system to offer unobtrusive health and wellness monitoring to support older adults living independently in their home{s}.  This involved the use of ambient sensors  non-intrusively {placed} around  the home to detect various {domestic, in-home} activities. The resulting time series data gathered from each  sensor can  be combined and analysed to create a model of an individual’s normal daily patterns of activity. 
During the first half of 2022, twenty-four healthy older adults aged 60 years {or older} and  living independently at home participated in a trial of the NEX project. The gender profile was predominantly female (81\% n=21) with a total population mean age of 73.2 years.  The majority of participants were independent and high functioning with only 8\% (n=2) reporting difficulties in completing activities of daily living (ADLS)\cite{katz1963studies} such as dressing etc. and only 4\% (n=1) reporting difficulties in completing more complex tasks defined as instrumental activities of daily living (IADLS) \cite{lawton1988instrumental} such as shopping for groceries etc.

A range of in-home sensors were installed in each participant's home including {(1)} contact sensors on the doors of cupboards or drawers {which were} used for kitchen  crockery, cutlery, delph, food staples and pots, {(2)} contact sensors on the fridge door and {(3)} electrical plug sensors on the kettle, microwave and toaster. There were also contact sensors on wardrobe door(s) and on drawer(s) in the bedroom, a 6-in-1 environmental sensor for temperature, humidity, light and presence in the bathroom and contact sensors on the front door, the patio door or the back door.  
{Collectively, data from t}hese {sensors} capture  the purposeful actions of the participant{s}, i.e. they indicate a deliberate action  thus {the} opening or closing of doors,  {the movement of a} person  and  {the} switching on or off of electrical devices. 
Sensor data was gathered from the in-home sensors from the 24 participants' homes over periods varying from 6 weeks to 6 months each  \cite{info:doi/10.2196/35277}.

The importance of the 24h circadian rhythm and the conformance that each of the NEX project participants has to that 24-hour rhythm, can be shown by computing the strength of 24-hour periodicity over all  data  from multiple sensor devices in the home.  The fusion of multiple data streams from multiple sensors captures the activities of daily living including food preparation and eating, self-grooming, cleaning, relaxing, and others. Periodicity intensity emphasises regularly and the regular recurrence of such activities in the home.

Periodicity intensity  {visualisations were}  presented to participants in the NEX project as a visualisation of their long-term behaviour and changes and the regularity of those trends over time. A periodicity intensity graph is a continuous timeline where high values indicate periods of regular behaviour as detected by the in-home sensors, and low values indicate periods of irregular activities. Such irregular activities are indicators of changes in the regular habits and can point to changes in their sleep patterns (sleep onset, duration, waking during the night, napping during the day), in  {their} movement  {into and out of, as well as around the home} (awake at night, sleeping during the day, staying in the home longer or shorter than normal), in self-grooming (using  {the} restroom at irregular times) and in eating (snacking during the day or during the night as opposed to  {having} regular meals).

For the NEX trial we used fused the data gathered from sensors in the homes of participants and computed a single overall periodicity intensity timeline for each participant and we present some of these in the Results section  {of this paper}.

\subsubsection{Periodicity Intensity in Student Learning Habits}

Many Universities have wellness programs to promote overall health and wellbeing of their staff and students \cite{baik2019universities}. Their benefits lead to improved health literacy, healthier mindsets, healthier relationships with others and positive changes to lifestyles regarding exercise, nutrition and {more}.  In our University we created and {run} an elective {undergraduate} course on wellness called FLOURISH{. This } also includes elements of data literacy by showing students how their own personal data, gathered from devices used to measure their sleep, exercise, nutrition, etc., {can} be used for good as part of students themselves monitoring these wellness factors. 

The wellness course was delivered entirely online via the University's virtual learning environment (VLE) Loop, a variation of Moodle and was accredited by the University, earning 5 ECTS for students who successfully completed it.  For each of the {10} wellness topics, the personal data that students were required to gather on themselves was not shared with the course instructor or with the University. However, as part of their participation in the course and to demonstrate how personal data gathered by the University as part of recording their online activities could also be used for good, each of the 169 students on the course authorised us to download their VLE access logs for all their taught courses for the semester. This included the {date, time and the URL of the Loop resource for each time that a student accessed any resource on the VLE}.

At the end of the teaching semester we used the log of all student accesses to all Loop resources on all their courses and generated a single periodicity intensity graph for each student covering all {online} accesses and shared the graph online privately  with them.  The resulting graphs provide{} an insight for each student into which period(s) of the preceding taught semester their learning habits as exemplified by their access to online resources, were more structured and regular.  This may help them in reflecting back on their learning throughout the semester{.}    Some sample graphs to illustrate this, as well as further insights, are presented in the Results section of this paper.

\subsubsection{Periodicity Intensity in Calf Movements}

The regularity of the sleep-wake cycle is important not just for humans but for other living animals. We  investigated the {strength of the} circadian periodicity of newborn calves on a commercial farm and how th{at} strength of  periodicity changes throughout their first 6–8 weeks after birth \cite{rhodes2022periodicity}.  Many animals including cattle are behaviourally synchronised with herds coordinating reactions to external zeitgebers \cite{conradt2003group} so our interest in this use case
is in the periodicity intensity of the circadian rhythm for the herd as a whole as well as for individual calves.

We gathered data from an Axivity AX3 wearable accelerometer \cite{de2021comparison} affixed to a collar worn on the necks of each of 24 dairy calves from just after their birth. The calves were fed twice a day, were bedded daily and had human contact during these activities. Feeding times were at similar times to milking as they were fed whole milk  thus their day-to-day activities would have been very structured and regular. Individual calf movements as detected by the accelerometers would have varied among calves {as} they play and interact and explore their surroundings so despite the regular structure to feeding and sleep, there is scope for a lot of individual variability among them. 

Data from the accelerometers was sampled at 12.5Hz and  gathered for up to  8 weeks for each calf.  It was processed in the way described in \cite{riaboff2019evaluation} by pre-processing  into signal vector magnitude (SVM), a time-series independent of  sensor orientation and thus invariant to  movement of the collar around the neck.  A Butterworth fourth-order band-pass filter \cite{fridolfsson2019effects} with frequency in the range 0.5 Hz {to} 20 Hz was then applied to remove any white noise, and negative values were converted into absolute values.  Movement values used for analysing periodicity were the aggregated mean of SVM calculated over non-overlapping 60-second windows.

\section{Results}

\subsection{Results of Periodicity Intensity in the Homes of Older Adults}

{We first examine and discuss the results of periodicity intensity calculation on sensor data from the homes of older citizens.}
The output from periodicity detection on data from {the} NEX project participants was calculated in real time and made available online.  These visualisations were used in debriefings with participants by their clinical carers and project researchers and supported by a short-form informative YouTube video to explain the graphs and to ensure the{ participants} understood how to interpret them.  

\begin{figure}[H]
\centering
\includegraphics[width=\linewidth]{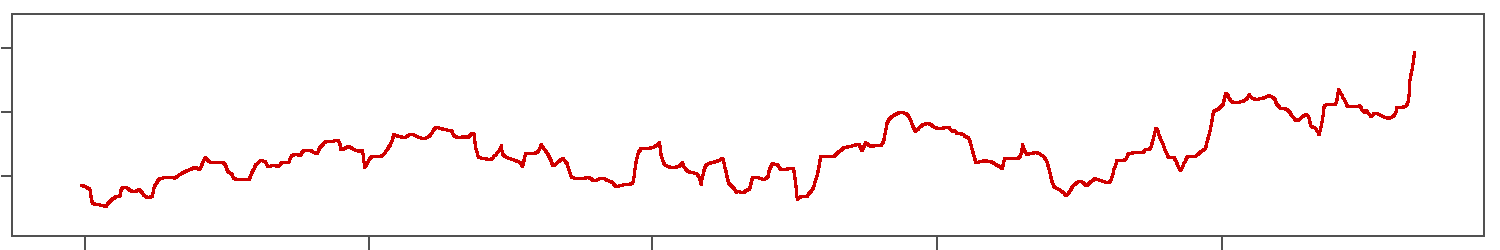}
\includegraphics[width=\linewidth]{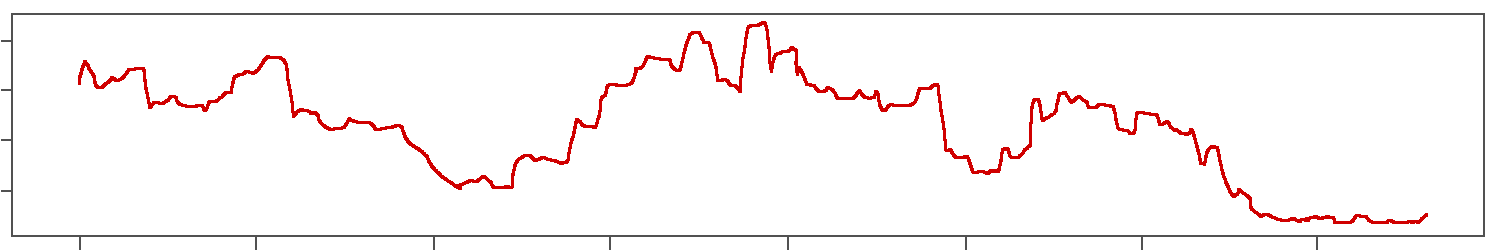}
\includegraphics[width=\linewidth]{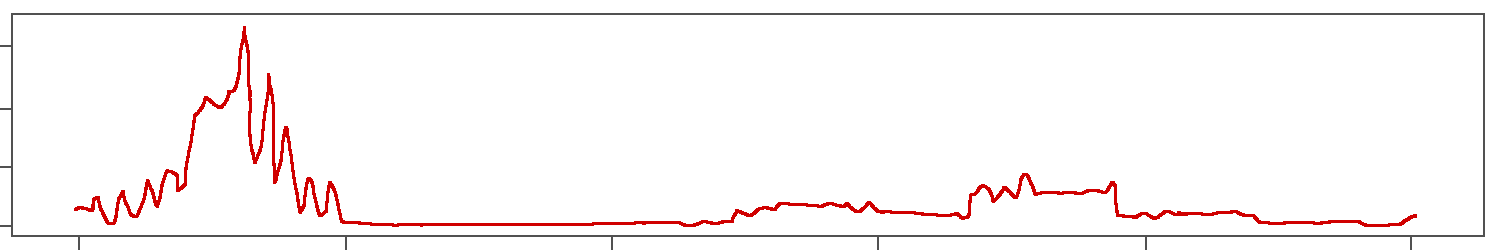}
\caption{{Normalised} periodicity intensity from 3 sample ARC trial participants for 3 months duration. Frequency is 24 hours, window size is 7 days and window shift is 1 hour. {X-axis values span 3 months, y-axis values have been normalised to the range 0 to 1.}\label{fig:NEX-participant}}
\end{figure}

Three example outputs are shown in Figure~\ref{fig:NEX-participant} where the values in each graph are normalised to a [0,1] range as actual periodicity intensity values will vary considerably across participants and each participant is interested in the relative changes in their own graph rather than in absolute values or {in} comparison to others.  Following experimentation to determine the best parameters to use, for each of the graphs the frequency for periodicity calculation is 24 hours reflecting interest in the circadian rhythm, the window is 7 days thus eliminating changes between weekends/weekdays and the overlap or stride length {} is 1 hour making the computation fast. Each graph represents a periodicity intensity over {a} 3 month{ period}.

The first graph in Figure~\ref{fig:NEX-participant} shows a participant with relatively stable and regular lifestyle and only small variations in periodicity intensity. The gradual rise in periodicity intensity towards the end of the 3 months shows a slight improvement in lifestyle regularity and there are no major perturbations throughout the 3 months. This participant is known to have a healthy, regular and comfortable lifestyle.

In the second graph in Figure~\ref{fig:NEX-participant} we see a participant with much more ups-and-downs in lifestyle. The increase in lifestyle regularity in the middle of the period corresponds to a constant presence at home due to catching Covid-19, self-isolating and recovering with regular behaviour and habits.  For this period the volume of sensor activations was much lower than in the earlier period as the participant was not doing much at home except spending time in bed, but it is not the number of sensor activations but their timing that contributes to periodicity intensity. Towards the end of the 3 month monitoring period this participant moved out of the home to recuperate with family {hence the flatlining at the end}.

The third graph in Figure~\ref{fig:NEX-participant} shows a participant who also caught COVID-19 during the early part of the 3 month recording period, and then soon afterwards also moved out of the{ir} home to recuperate with family which was indicated by the flatlining in the graph. The slight rise in periodicity intensity during the second half of the 3 months indicates family calling to the home on semi-regular basis to collect mail and check it for security. 

The result of this analysis is a form of self-monitoring {of the individual} but with deeper insights into overall behaviour and activities. This may be used to promote or sustain long-term behaviour change as the subject seeks to maintain
high levels of rhythmicity in their lives.

\subsection{Results of Periodicity Intensity in Student Learning Habits}

{We now present an in-depth analysis and discussion of the results of applying periodicity intensity to students' online learning habits.}  The periodicity intensity graphs for each student were made available to them at end of the teaching semester, 
supported by a short-form informative YouTube video to explain the graphs and to ensure they understood how to interpret them.
Five  sample graphs are shown in Figure~\ref{fig:Student-participants}. In each graph the periodicity intensity values are normalised to a [0,1] range to smooth out variance across students and to make each graph equally legible. {This is } similar to the older adult participants in the previous study {{in that} each student will be interested in the relative changes in their own graph rather than in absolute values or {in} comparison to others.  We also added 3 coloured vertical bands to each graph, red indicating the period of lowest periodicity intensity, green indicating the period of highest and yellow indicating the period of greatest (steepest) change. The parameters used in the calculation of the graphs {are} a frequency of 24 hours, window size is 7 days and window shift or stride is 3 hours.

\begin{figure}[H]
\centering
\includegraphics[width=\linewidth,trim={0 0 51mm 9mm},clip]{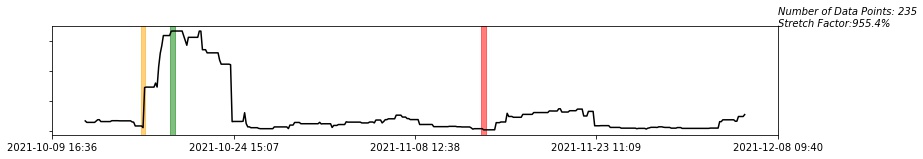}
\includegraphics[width=\linewidth,trim={0 0 51mm 9mm},clip]{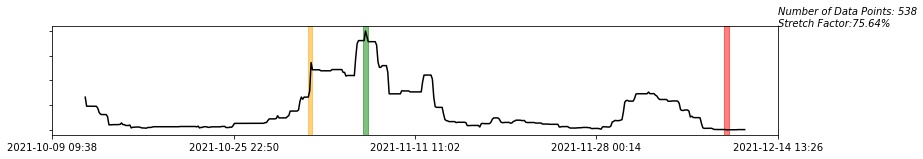}
\includegraphics[width=\linewidth,trim={0 0 51mm 9mm},clip]{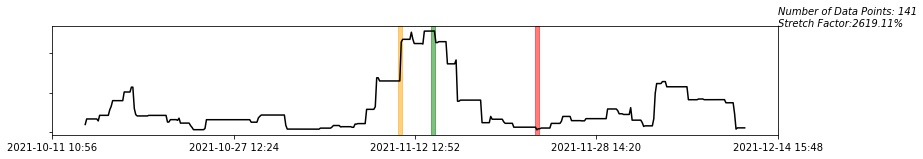}
\includegraphics[width=\linewidth,trim={0 0 51mm 9mm},clip]{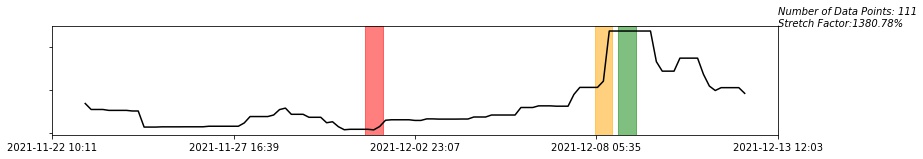}
\includegraphics[width=\linewidth,trim={0 0 47.5mm 9mm},clip]{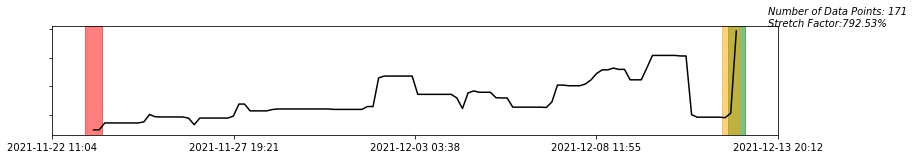}
\caption{{Normalised} periodicity intensity for a sample of five student participants. Frequency
is 24 hours, window size is 7 days and window shift or stride is 3 hours. {Y-axis values have been normalised to the range 0 to 1.}\label{fig:Student-participants}}
\end{figure}   

The five graphs shown in Figure~\ref{fig:Student-participants} have  different start dates reflecting the different times students started studying and they reflect totally different study patterns with the first, second, third and fourth students peaking at different times during the semester in terms of the structured and regular nature of their online access. The first {student has her/his peaks}  earliest {in the semester while}  the fourth {student has her/his peaks}  last. The graph for the fifth student shows a student with a gradual increase in regularly of studying throughout the semester with a burst of regular studying at the end, just before the course examinations.

While individual periodicity intensity graphs are of value for each student we can get further value from this analysis by time-aligning and stacking the 169 individual student graphs as shown in Figure~\ref{fig:All-student-participants}.
For these we set a fixed start and end date/time before {stacking} the graphs. Since periodicity calculations are for windows of 7-days, there is a gap of 3.5 days at the beginning and the end of the graph. 

\begin{figure}[H]
\centering
\includegraphics[width=\linewidth,height=0.25\textheight]{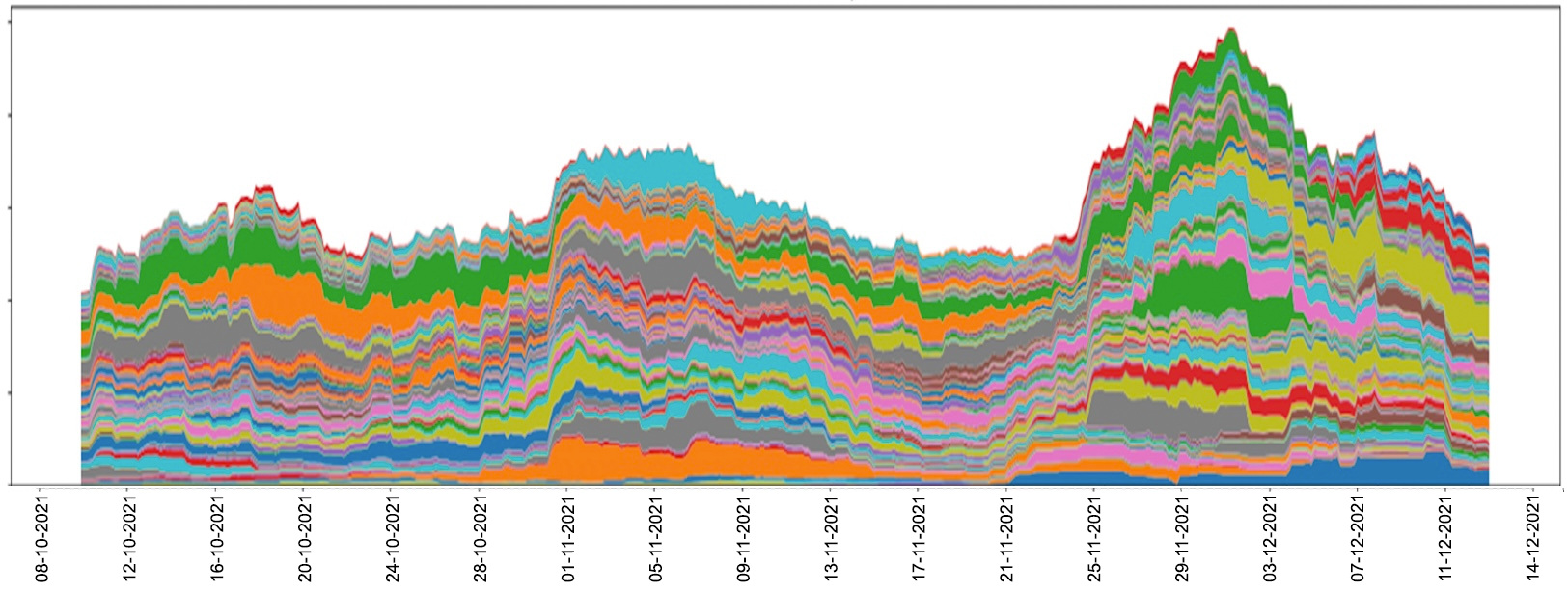}
\caption{Stacked line chart of cumulative periodicity intensity from all 169 student participants. {Y-axis values have been normalised to the range 0 to 1.}\label{fig:All-student-participants}}
\end{figure}   

Figure~\ref{fig:All-student-participants} shows the aggregated periodicity intensities for the  cohort of 169 students where the y-axis is a stack of periodicity intensities for each student, each student shown in a different colour. The top line of the overall graph show the aggregated periodicities from all students and it reveals three distinct peaks corresponding to the early part of semester as students settle into University life, the middle of semester as students take advantage of the mid-semester reading week to catch up on study, and at the end of semester, just before examinations, as students prepare for those examinations.

As shown by \citeauthor{byron2008stacked} in \cite{byron2008stacked}, stacked line graphs may present an illusion of peaks and troughs which might be caused by  a subset of students rather than {the peaks and troughs appearing} across all {students}. Thus the overall stacked line graph might be artificially boosted or deflated by peaks or troughs from a small number of students. To assess this, we divided the students into random sub-groups and plotted a stacked line graph for each sub-group. Since the stacked line graphs for the sub-groups show the same shape as the stacked line graph for the whole cohort there {are} no artificial boosts or troughs from a sub-group {of students}, and the pattern is generalised across the cohort.

We can see from Figure~\ref{fig:All-student-participants} that there is a fault line just before 1st November. The University's academic calendar shows this was the date for  closing of registrations for the student course choices.  A second interesting fault line appears just before 25th November. The semester's teaching period  ended on 27th November and the pre-examination study period  was between 29th November and 5th December. We can see  an increase in the structured nature of VLE activities  from  students with their assignment submissions and {revision} of study materials in preparation for examinations, and this tails off as each student finishes their examinations.

The result of this analysis of periodicity intensity is personalised feedback for each student on their learning habits with deeper insights into the learning behaviour and activities of the overall student cohort. This has personal benefit for each student as well as benefit for the University and course directors.

\subsection{Results of Periodicity Intensity in Calf Movements}

{Our final in-depth discussion is of the periodicity intensity algorithm applied to accelerometer data from new-born calves.}
The 24 calves used in the calf movement study had different dates of birth and {their} sensor data thus started  on different dates. For the oldest calf (sensor ID 20714)  the accelerometer data began on 2nd March at 09:00 and for the youngest calf the accelerometer data began on  5th April at 10:00. Though  start dates varied, all  calves in the herd were contributing accelerometer data by 5th April thus {it is} from that date onwards that we  obtain a  picture of the full herd’s periodicity.  Of the 24 calves, 5 are considered as outliers and  did not contribute to the analysis of the overall herd because their collars had fallen off at some point during the data logging  and if reattached it would have created a gap in the data which would have impacted the calculation of periodicity intensity for a period of more than 2 weeks. The 19 remaining calves were used in the analysis. The graphs of periodicity intensity for individual calves can be seen in Figure~\ref{fig:Individual-calves} and {their x-axes} are time-aligned.  For these the periodicity frequency was set to 24 hours, the window size was 7 days and the stride {length} or shift of the time-lagged overlapping windows was 15 minutes.

\begin{figure}[H]
\centering
\includegraphics[width=\linewidth]{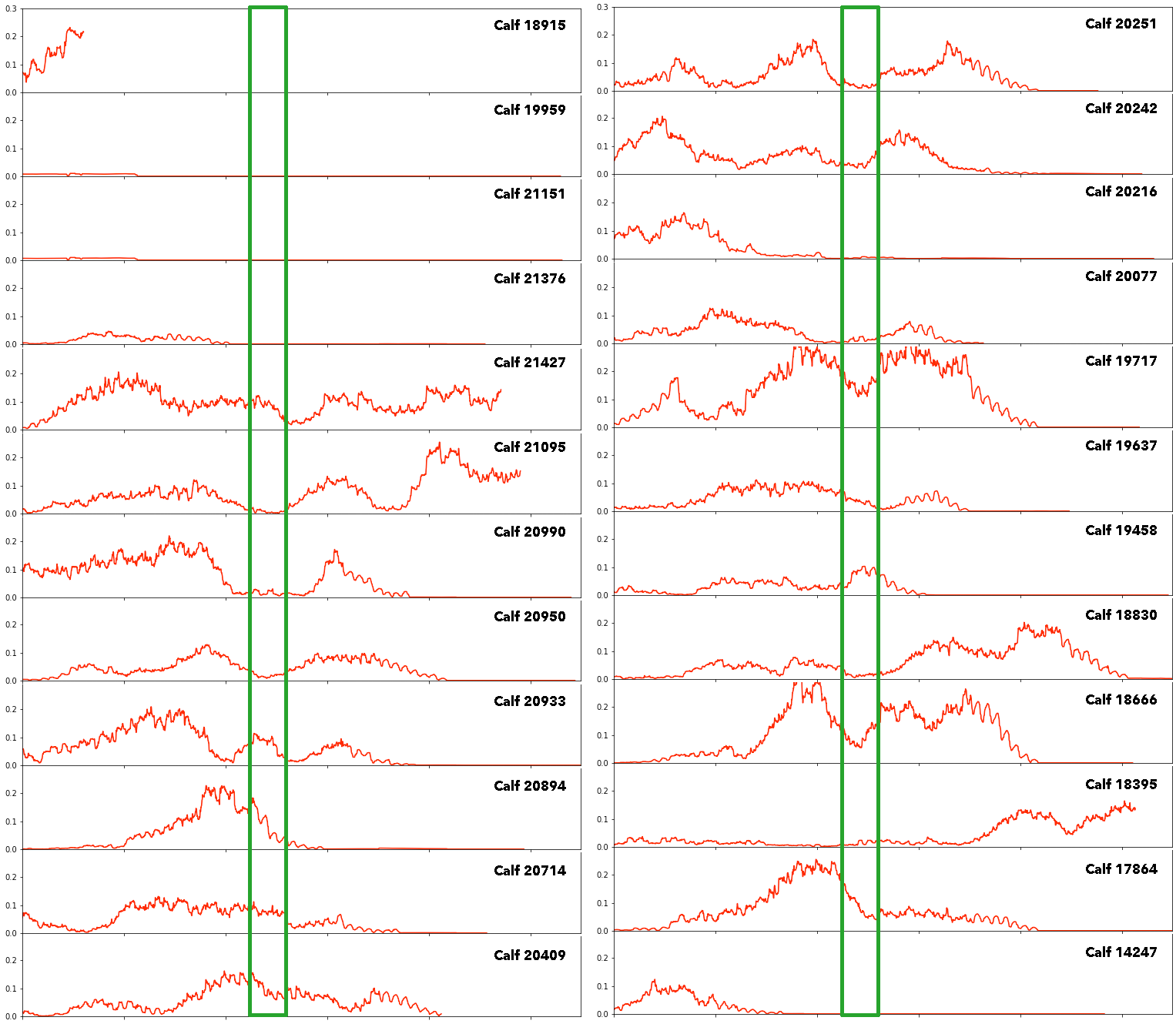}
\caption{Line graphs of calf periodicity intensities for each of 24 calves.
Frequency is 24 hours,
window size is 7 days and window shift is 15 minutes. The vertical green lines indicates the dates of disbudding. \label{fig:Individual-calves}}
\end{figure}   

Figure~\ref{fig:Individual-calves} shows the 24 individual periodicity intensity graphs and the first 4 in the left column and the 3rd in the right column were discarded for reasons mentioned earlier. The vertical green lines in the graph indicate the dates of disbudding, a process of surgical removal of horn buds before the {calves} grow and cause a danger to other calves and to the farmer \cite{espinoza2020effect}.  

Looking across the set of 19 graphs it is difficult to discern any coordinated pattern across the herd.  When we generate a stacked  graph of the {time-aligned} periodicity intensities in the same way as was done for the student VLE accesses, we get the graph shown in Figure~\ref{fig:All-calves}, where each colour represents periodicity intensity for a different calf. 

\begin{figure}[H]
\centering
\includegraphics[width=\linewidth,height=0.3\textheight]{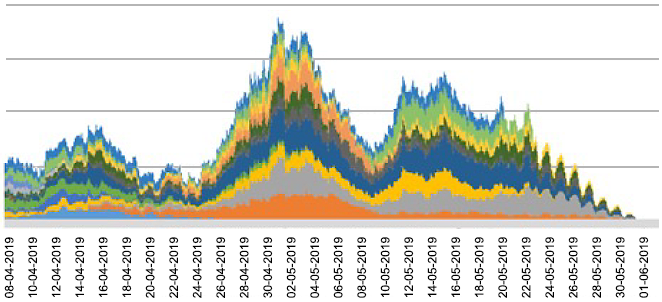}
\caption{Stacked line chart of cumulative calf periodicity intensity from 19 of the 24 calves, from \cite{rhodes2022periodicity}. {Y-axis values have been normalised to the range 0 to 1.}\label{fig:All-calves}}
\end{figure}   

The staked line graph of periodicity intensities shows a  peak at around the beginning of May followed by a clear drop in 24h periodicity intensity across the herd, indicating a synchronous dis-improvement in herd welfare. This reached its low point after about 1 week, around 9 May and was followed by a return to better herd welfare, although not as high  at {around} 2 May,  Because of the 7-day windowing in our calculations the drop-off, which reached its minimum point around 9 May, was most likely caused by a stressful or traumatic event at around 2 May which is the time window during which the disbudding process took place.

As with the use case of student learning habits from a VLE, we divided the calves into random sub-groups and plotted a stacked line graph for each sub-group. These show the same shape as the stacked line graph for the whole herd thus there is no artificial boosts or troughs from a sub-group of calves, and the pattern is generalised across the herd.

The result of this analysis of periodicity intensity is similar to that from the analysis of student accesses to their VLE.  We have individual feedback on a wellbeing indicator for each calf based on their movements with deeper insights into the wellbeing of the overall herd and this is of interest to the farmer.

\section{Discussion}

This paper presents three applications of   calculating  periodicity intensity throughout a time series {on} three {very} different real world use cases. In each case we gain deeper insights into the data than would have been revealed without periodicity analysis. These powerful insights into {the underling} time series data can be obtained from any data source that has {underlying} periodicity {as an inherent feature}.

The value of this form of data analysis is that it is unsupervised, requires no training data  and thus it is generalisable.  {This makes the value of the technique in terms of economic impact quite attractive as it can be applied to any use case where there is longitudinal time series data to be analysed in order to detect gradual shifts or changes over time.}

{In terms of implementation, the choice of values for the three parameters are not difficult to determine.} 
The frequency of the recurring pattern to be investigated, $Fo$, is often 24 hours but not always so and the use case usually already knows what this should be. The size of the window in which the periodicity is calculated $m$ is also usually known in advance. For example if $Fo$ is 24 hours then setting $m$ to 7 days nullifies any effects of weekend/weekday behaviour transitions in the underlying data. The overlap between windows or stride {length} $l$ is influenced by the demands of the use case and the amount of computation resources available and not any algorithmic limitation. The technique is agile in terms of the number and even the type of time series data used, and it is fast to compute.  Because we use the Lomb-Scargle algorithm \cite{VanderPlas_2018} to calculate periodicity in windows it handles irregularly sampled data such as our student accesses to their VLE, as well as moderate amounts of missing data, {though} not completely missing data as in those calves who's collars fell off but were replaced quickly in our third use case.  {Finally, the algorithm for calculating periodicity intensity is mathematically deterministic, producing the same output from the same input repeatedly.}
 
Periodicity intensity is a complex metric to fully comprehend because it is novel and non-conventional. In the two of our use cases which involved human participants -- in-home sensors in the homes of older adults and students accessing their VLE -- we needed the support of an animated helper video to describe what the periodicity intensity was because our users were healthcare professionals and older adults.

For future work we will concentrate on ways to make the metric more intuitively understandable by mapping changes in a periodicity intensity graph back to the underlying data thus showing exactly what data has caused changes in periodicity intensity and this would be useful in those use cases involving human subjects. We will also explore the robustness of the algorithm as a function of data quality by introducing noise to see how a decrease in data quality impacts the generated output.  Finally we will focus the implementation of the Lomb-Scargle algorithm to determine periodicity at the 24 hour frequency only as our experience has been that this frequency setting is the one that most use cases are interested  in.


\vspace{6pt} 



\authorcontributions{Conceptualization,  A.S. and F.H.; 
methodology, A.S. and F.H.; 
software, F.H.; 
validation,  A.S. and F.H.; 
formal analysis, A.S. and F.H.; 
investigation, A.S. and F.H.; 
data curation, A.S.; 
writing---original draft preparation,  A.S.; 
writing---review and editing,  A.S and F.H..
All authors have read and agreed to the published version of the manuscript.
}

\funding{This research was funded by Science Foundation Ireland grant number SFI/12/RC/2289\_P2, cofunded by the European Regional Development Fund and by the Disruptive Technologies Innovation Fund administered by Enterprise Ireland, project grant number DT-2018-0258 and by a UCD Wellcome Institutional Strategic Support Fund which was financed jointly by University College Dublin and the SFI-HRB-Wellcome Biomedical Research Partnership (ref 204844/Z/16/Z).}

\institutionalreview{The study was conducted in accordance with the Declaration of Helsinki, and ethical approvals for (1) the study involving sensor data from the homes of older adults was obtained from the DCU Research Ethics Committee  (DCUREC202221) and (2) the study involving log data from student access to an online VLE was approved by DCU’s Data Protection Unit 07042021-DPO with ethics approval was given by the School of Computing Research Ethics Committee.  As ethical approval for (2)  was deemed to be notification-only, the institutional policy is that only School-level ethics approval is needed.
The animal study protocol for the study involving sensor data from movement and behaviour of the herd of cattle were collected under an ethical exemption from UCD Animal Research Ethics Committee, approval number AREC-E-19-46-McAloon.
}

\informedconsent{Informed consent was obtained from all human subjects involved in the studies.}

\dataavailability{
The data presented in  the study involving sensor data from the homes of older adults are openly available in FigShare at \url{https://doi.org/10.6084/m9.figshare.21415836.v2}.
The data presented in the study involving student access to an online VLE are openly available in Figshare at \url{https://doi.org/10.6084/m9.figshare.20288763.v2}.
The data presented in the study involving movement and behaviour of a herd of cattle are openly available in Figshare at \url{https://doi.org/10.6084/m9.figshare.20039486.v1}.}

\conflictsofinterest{The authors declare no conflict of interest. The funders had no role in the design of the study; in the collection, analyses, or interpretation of data; in the writing of the manuscript; or in the decision to publish the~results.} 



\abbreviations{Abbreviations}{
The following abbreviations are used in this manuscript:\\

\noindent 
\begin{tabular}{@{}ll}
ADL & Activity of daily living\\
ECTS & European credit transfer and accumulation system\\
IADL & Instrumental activity of daily living\\
RP & Recurrence plots\\
SVM & Signal vector magnitude\\
VLE & Virtual learning environment\\
{EMD} & {Empirical Mode Decomposition }\\
{ZCR} & {Zero crossing rate }\\
{ZCD} & {Zerocross Density Decomposition }\\
\end{tabular}
}




\begin{adjustwidth}{-\extralength}{0cm}

\reftitle{References}


\bibliography{references}

\end{adjustwidth}
\end{document}